\documentclass[12pt]{article}
\usepackage[english]{babel}
\usepackage{ulem}
\usepackage[utf8]{inputenc}
\usepackage[T1]{fontenc}

\def\bPi{\mathbf{\Pi}}

\def\mK{\mathcal{K}}

\usepackage{amsmath}

\def\mC{\mathcal{C}}
\usepackage{amsfonts}

\usepackage{url}
\usepackage[dvips]{graphicx}
\usepackage{calc}
\usepackage{subfigure}
\usepackage{multirow}

\def\mL{\mathcal{L}}
\def\mH{\mathcal{H}}

\def\bA{\mathbf{A}}

\def\bAi{\left(\bA^{-1}\right)}

\begin{document}
	\begin{titlepage}
	\begin{center}
		{\Large{ \bf Note About  Born-Infeld Inspired Gravity Coupled Non-Minimally  		to Scalar Fields
and Eddington Gravity with Matter	 }}
		
		\vspace{1em}  
		
		\vspace{1em} J. Kluso\v{n} 			
		\footnote{Email addresses:
			klu@physics.muni.cz (J.
			Kluso\v{n}) }\\
		\vspace{1em}
		\textit{Department of Theoretical Physics and
			Astrophysics, Faculty of Science,\\
			Masaryk University, Kotl\'a\v{r}sk\'a 2, 611 37, Brno, Czech Republic}
		
		\vskip 0.8cm
		
		%
		%
		%
		%
		%
		%
		
		\vskip 0.8cm
		
	\end{center}

\begin{abstract}
	We study canonical formulation of  Born-Infeld inspired gravity coupled non-minimally to  scalar field.  Then we propose form of Eddington Gravity coupled to collection 
	of scalar fields whose canonical form is the same as Hamiltonian for General Relativity coupled minimally to scalar field. 
	\end{abstract}

\bigskip

\end{titlepage}

\newpage

\section{Introduction and Summary}\label{first}
Recently we studied canonical formulation of Eddington and Born-Infeld inspired gravities
\footnote{For review of Born-Infeld inspired gravities and extensive list of references, see
\cite{BeltranJimenez:2017doy}.} in 
\cite{Kluson:2025ajf,Kluson:2025uyl}. The main result derived in these papers is that the dynamical degrees of freedom of gravity corresponds to the momentum conjugate to the specific combination of connection 
introduced in \cite{Horava:1990ba}. In more details, we studied canonical form of these specific models with the help of Faddeev-Jackiw treatment of constraints systems 
\cite{Faddeev:1982id,Faddeev:1988qp,Jackiw:1993in}. We identified true physical degrees of freedom and we showed that in case of Eddington gravity its canonical form is equivalent to the canonical form of General Relativity action \cite{Dirac:1958sc,Arnowitt:1962hi,Gourgoulhon:2007ue}. In case of Born-Infeld gravity which minimally couples to matter the situation is more complicated since the action now depends on 
auxiliary metric \cite{Kluson:2025uyl} which however is not the same as the physical one. On the other hand this  non-dynamical metric has to be integrated out resulting into Hamiltonian constraint that differs from the Hamiltonian constraint of Einstein-Hilbert action coupled minimally to matter in a significant way. This is expected result since Born-Infeld inspired gravity contains new mass scale that 
determines when this theory differs from the General Relativity action. However the  original motivation for formulation of Eddington gravity was to have theory with the first order of derivatives 
\cite{Eddington} which is equivalent to General Relativity in case of absence of matter. On the other hand it is not completely clear how to include matter   content to Eddington gravity
(Nice discussion can be found in  \cite{Padmanabhan:2013nxa}) so that one can ask the question whether some form of Born-Infeld inspired gravity that non-minimally couples to matter could give milder modification of General Relativity action. Such a form of Born-Infeld gravity was proposed in \cite{Vollick:2005gc}
and our goal is to perform its canonical analysis and to explicitly show how it is related to the canonical form of General Relativity at least in case when the matter is represented by collection of scalar fields. 
This analysis is based on our previous papers \cite{Kluson:2025ajf,Kluson:2025uyl} where the main difference is now in presence of scalar fields in Lagrangian so that we can skip many details which can be found in 
\cite{Kluson:2025ajf,Kluson:2025uyl}. We find canonical form of the action that differs from the General Relativity action  by presence of the term proportional to $V^2$ where $V$ is scalar field potential. Moreover, if we analyze modified action which was suggested in \cite{Vollick:2005gc} we find that the resulting canonical form has exactly the same form as General Relativity action coupled minimally to scalar fields. An important
outcome of this analysis is an observation, that space-time metric which appears in Born-Infeld inspired gravity, is auxiliary field that is eliminated from the action by solving corresponding equations of motion. Motivated by this observation we propose form of Eddington gravity that couples to the collection of scalar fields without any metric dependence. Then when we perform canonical analysis of this action and we find that dynamical metric degrees of freedom arise naturally as canonical momenta. We further show that  resulting canonical form of action is completely equivalent to the General Relativity and scalar field action. 
We mean that it is very nice result  that we are able to construct Eddington action that couples to matter in the form of scalar field.

Natural extension of this work is to include more general form of matter. Such possibility was discussed in 
\cite{Vollick:2005gc} where matter contend in the form of electromagnetic and spinor fields was analyzed. However it was also shown there \cite{Vollick:2005gc} that resulting equations of motion cannot be solved in an exact form. On the other hand
it would be nice to find canonical form of Born-Infeld inspired gravity that couples to electromagnetic or spinor field. 
 We hope to return to these problems in future. 

This paper is organized as follows. In the next section (\ref{second}) we introduce Born-Infeld inspired gravity that non-minimally couples to matter and find its canonical form. Then in section (\ref{third}) we propose Eddington gravity that couples to scalar field and we  again determine its canonical form.

\section{Hamiltonian Formalism for Born-Infeld Inspired Gravity Non-minimally Coupled to Scalar Fields }\label{second}
In this section we introduce an action for Born-Infeld inspired gravity that couples non-minimally to collection of scalar fields.
Such an action  was proposed in 
\cite{Vollick:2005gc}
 and  has the form 
\begin{eqnarray}\label{SBIM}
	&&
	S_{BIM}=M_p^2M_{BI}^2\int d^4x [\sqrt{-\det \bA}-\lambda\sqrt{-g}] \ , \nonumber 
	\\
	&&
	\bA_{\mu\nu}=g_{\mu\nu}+\frac{1}{M^2_{BI}}(R_{(\mu\nu)}+\frac{1}{M^2_p}
(\partial_\mu\phi^A \partial_\nu\phi^BK_{AB}+\frac{1}{4}g_{\mu\nu}V (\phi^A) )	\ , 
	\nonumber \\
&& 	R_{(\mu\nu)}=\frac{1}{2}(R_{\mu\nu}+R_{\nu\mu}) \  
, \quad R_{\mu\nu}=\partial_\gamma\Gamma^\gamma_{\mu\nu}-\partial_\mu \Gamma^\gamma_{\nu\gamma}+\Gamma^\gamma_{\gamma \alpha}
\Gamma^\alpha_{\mu\nu}-\Gamma^\gamma_{\mu \beta}\Gamma^\beta_{\gamma \nu} \ , \nonumber \\	
\end{eqnarray}
where $M_{BI}$ is the mass scale that determines when higher curvature corrections are important. Further, $\phi^A, A=1,\dots,K$ is set  of scalar fields $\phi^A$ with potential 
$V(\phi)$ and non-degenerative kinetic matrix $K_{AB}$. Finally $\lambda$ is constant and multiplies determinant of four dimensional metric $g_{\mu\nu} \ , \mu,\nu=0,1,2,3$ and $\Gamma^\gamma_{\mu\nu}=\Gamma^\gamma_{\nu\mu}$ are coefficients of connection which are true dynamical degrees of freedom of gravity. It is important to stress that there is no relation between $g_{\mu\nu}$ and $\Gamma^\rho_{\mu\nu}$ in Born-Infeld
inspired gravity.

As a check of the consistency of this proposed action note that  in the limit of large $M_{BI}$ we can write the action (\ref{SBIM}) as
\begin{eqnarray}
&&S_{BIM}=M_p^2 M_{BI}^2(1-\lambda)\int d^4x \sqrt{-g}+
\frac{M^2_p}{2}\int d^4x \sqrt{-g}g^{\mu\nu}R_{\mu\nu}(\Gamma)+\nonumber \\
&&+
\frac{1}{2}\int d^4x \sqrt{-g}(g^{\mu\nu}\partial_\mu\phi^A
\partial_\nu\phi^B K_{AB}+V) \  
\nonumber \\
\end{eqnarray}
that corresponds to the General Relativity action in Palatini formulation  with cosmological constant $\Lambda=\frac{1}{2}M_{BI}^2M^2_p(\lambda-1)$ minimally coupled to scalar fields.

Our goal is to find canonical form of the action (\ref{SBIM}) in order to identify true physical degrees of freedom. We follow our analysis performed in previous papers 
\cite{Kluson:2025ajf,Kluson:2025uyl}.

It is convenient to express symmetric part of Ricci tensor in $d-$space time dimensions with the help of variable 
$G_{\mu\nu}^\lambda$ defined
as \cite{Horava:1990ba}
\begin{equation}
	G_{\mu\nu}^\lambda=
	\Gamma_{\mu\nu}^\lambda-\frac{1}{2}(\delta_\mu^\lambda
	\Gamma^\rho_{\rho\nu}+\delta_\nu^\lambda \Gamma^\rho_{\rho\mu})
\end{equation}
so that
\begin{equation}
	R_{(\mu\nu)}=\frac{1}{2}(R_{\mu\nu}+R_{\nu\mu})=
	\partial_\lambda G^\lambda_{\mu\nu}+
	\frac{1}{d-1}G^\lambda_{\lambda\mu}G^\sigma_{\sigma\nu}
	-G^\lambda_{\sigma\mu}G^\sigma_{\lambda\nu}	\ . 
\end{equation}
Let us use these variables for calculations of conjugate momenta
\begin{eqnarray}\label{defpi}
	&&	\Pi^{\mu\nu}_0\equiv \bPi^{\mu\nu}=\frac{\partial \mL}{\partial (\partial_0 G ^0_{\mu\nu})}=
	\frac{M_p^2}{2}\sqrt{-\det \bA}\bAi^{\mu\nu} \  ,  \nonumber \\
	&&	\Pi^{\mu\nu}_i=\frac{\partial \mL}{\partial (\partial_0 G^i_{\mu\nu})}=0  \ , 
	\nonumber \\
	&&	p_A=\frac{\partial \mL}{\partial(\partial_0\phi^A)}
=\sqrt{-\det \bA}K_{AB}\partial_\nu \phi^B \bAi^{\nu 0} \ , \nonumber \\
\end{eqnarray}
where  $\bAi^{\mu\nu}$ is matrix inverse to $\bA_{\mu\nu}$ 
\begin{equation}
	\bA_{\mu\rho}\bAi^{\rho\nu}=\delta_\mu^\nu \ . 
\end{equation}
Now using (\ref{defpi})
 we obtain following Hamiltonian 
\begin{eqnarray}
&&\mH=\bPi^{\mu\nu}\partial_0 G^0_{\mu\nu}+p_A\partial_0\phi^A-\mL=\nonumber \\
&&=M_p^2M_{BI}^2\sqrt{-\det \bA}+M_p^2M_{BI}^2\lambda\sqrt{-\det g}-
\nonumber \\
&&-M^2_{BI}\bPi^{\mu\nu}g_{\mu\nu}-
\bPi^{\mu\nu}(\partial_iG^i_{\mu\nu}+\frac{1}{3}G^\lambda_{\lambda\mu}
G^\sigma_{\sigma\nu}-G^\lambda_{\sigma\mu}G^\sigma_{\lambda\nu}) \nonumber \\
&&+\frac{1}{2}\sqrt{-\det \bA}\bAi^{0\nu}
\partial_\nu\phi^A \partial_0 \phi^B K_{AB}
-\frac{1}{2}
\sqrt{-\det \bA}\bAi^{i\nu}
\partial_\nu\phi^A \partial_i \phi^B K_{AB}-\nonumber \\
&&
-\frac{1}{4M_p^2}\bPi^{\mu\nu}g_{\mu\nu}V \ .
\nonumber \\
\end{eqnarray}
Finally we have to express Hamiltonian density given above as function of canonical variables. First of all from definition of 
$\bPi^{\mu\nu}$ given on the first line in (\ref{defpi}) we find 
\begin{equation}
	\det \bA=\bPi (\frac{2}{M_p^2})^4 \ , \quad \bPi\equiv \det \bPi^{\mu\nu}
\end{equation}
so that 
\begin{equation}
	\bAi^{\mu\nu}=\frac{M_p^2}{2}\frac{1}{\sqrt{-\bPi}}\bPi^{\mu\nu} \ . 
\end{equation}
Then from the definition of momenta $p_A$ given on the second line
in (\ref{defpi}) we get
\begin{equation}
	\partial_0\phi^A
=	\frac{M_p^2}{2\bPi^{00}}K^{AB}p_B-\partial_i\phi^A\frac{\bPi^{i0}}{\bPi^{00}} \ . 
\end{equation}
Then after some calculations we obtain final form of Hamiltonian density for the action (\ref{SBIM}) 
\begin{eqnarray}
&&\mH=-M^2_{BI}\bPi^{\mu\nu}g_{\mu\nu}-\bPi^{\mu\nu}
(\partial_iG^i_{\mu\nu}+\frac{1}{3}G^\lambda_{\lambda \mu}G^\sigma_{\sigma\nu}-G^\lambda_{\sigma\mu}G^\sigma_{\lambda\nu})- \nonumber \\
&&-\frac{1}{4M^2_p}\bPi^{\mu\nu}g_{\mu\nu}V+4\frac{M^2_B}{M^2_p}\sqrt{-\bPi}+M^2_pM^2_{BI}\lambda\sqrt{-\det g}
\nonumber \\
&&+\frac{M_p^2}{4\bPi^{00}}p_AK^{AB}p_B-\frac{\bPi^{i0}}{
\bPi^{00}}p_A\partial_i\phi^A-
\frac{1}{M_p^2\bPi^{00}}(\bPi^{00}\bPi^{ij}-\bPi^{0i}\bPi^{0j})
K_{AB}\partial_i\phi^A\partial_j\phi^B \ . 
\nonumber \\
\end{eqnarray}
An important property of this canonical formalism is that 
$G_{\mu\nu}^i$ together with $g_{\mu\nu}$ are non-dynamical.
Now we can proceed as in our previous papers 
\cite{Kluson:2025ajf,Kluson:2025uyl} where we 
used analysis introduced in  \cite{Faddeev:1982id}
which  is based on alternative procedure of constraints systems \cite{Faddeev:1988qp,Jackiw:1993in}
\footnote{Standard canonical analysis of first order gravities
	was performed in 
	\cite{McKeon:2010nf,Chishtie:2013fna,Kiriushcheva:2011aa,Kiriushcheva:2010ycc,Kiriushcheva:2010pia,Kiriushcheva:2006gp}
	where conjugate momenta to $g_{\mu\nu}$ and $G_{\mu\nu}^i$ are  primary constraints and then it is checked their stability during the time evolution of system.}. In this approach 
we solve equations of motion for non-dynamical variables and their solutions we insert back  to the action.
 We start with the equations of motion for  $G_{\mu\nu}^i$ 
\begin{eqnarray}
	\Sigma_i^{\mu\nu}
	=\partial_i\bPi^{\mu\nu}+\Gamma^{\mu}_{i\beta}\bPi^{\beta\nu}+
	\Gamma^\nu_{i\beta}\bPi^{\beta\mu}-\Gamma^\sigma_{\sigma i}
	\bPi^{\mu\nu}= 0 \nonumber \\
\end{eqnarray}
which has the form of covariant derivative of $\bPi^{\mu\nu}$ where the last term is a consequence of the fact that 
$\bPi^{\mu\nu}$ is tensor density. 
Then  solving the equation $\Sigma_i^{00}=0$ we get
\begin{eqnarray}
	\Gamma^0_{i0}
	=-\frac{1}{\bPi^{00}}(\nabla_i \bPi^{00}+\Gamma^0_{im}\bPi^{m0}) \ , \nonumber \\ 
\end{eqnarray}
where we introduced spatial covariant derivative 
\begin{equation}
	\nabla_i\bPi^{00}=\partial_i \bPi^{00}-\gamma^m_{mi}\bPi^{00} \ , 
\end{equation}
where we used the fact that $\bPi^{00}$ is scalar density and where coefficients of connection $\gamma^i_{jk}$ are defined as
\begin{equation}
	\gamma^i_{jk}=\Gamma^i_{jk}-\frac{\bPi^{i0}}{\bPi^{00}}\Gamma^0_{jk} \ . 
\end{equation}
Further, solving equations $\Sigma^{0i}_j=0$ we obtain 
\begin{eqnarray}
	\Gamma^j_{0i}=-\frac{1}{\bPi^{00}}
	(\nabla_i\bPi^{0j}+\Gamma^0_{im}\bPi^{mj}) \ ,  \nonumber \\
\end{eqnarray}
where
\begin{equation}
	\nabla_i\bPi^{0j}=\partial_i\bPi^{0j}+\gamma^j_{ik}\bPi^{k0}-
	\gamma^m_{mi}\bPi^{0j} \ . 
\end{equation}
Then we obtain  canonical  action in the form 
\begin{eqnarray}
&&S=\int d^4x (\partial_t q^{ij}\mK_{ij}+\partial_t\phi^A p_A-
\frac{1}{\bPi^{00}}\mC-\frac{\bPi^{0i}}{\bPi^{00}}(
\mC_i-p_A\partial_i\phi^A)+\nonumber \\
&&+\bPi^{\mu\nu}M^2_{BI}g_{\mu\nu}-
\lambda M^2_{BI}M^2_p\sqrt{-\det g}+\frac{1}{4M_p^2}\bPi^{\mu\nu}g_{\mu\nu}V) \ , \nonumber \\
\end{eqnarray}
where we defined $q^{ij}$ and their conjugate momenta $\mK_{ij}$  as
\begin{equation}
	q^{ij}=\bPi^{0i}\bPi^{0j}-\bPi^{00}\bPi^{ij} \ , 
	\quad \mK_{ij}=\frac{1}{\bPi^{00}}\Gamma^0_{ij} \ , 
\end{equation}
and 
where $\mC$ and $\mC_i$ are defined as
\begin{eqnarray}
	&&	\mC=4\frac{M_{BI}^2}{M^2_p}\sqrt{\det q^{ij}}-\mK_{ij}q^{im}q^{jn}\mK_{mn}+
	\mK_{ij}q^{ij}\mK_{mn}q^{mn}-\nonumber \\
	&&	- q^{nm}\partial_n\gamma^p_{pm}+
	q^{ij}\partial_m\gamma^m_{ij}-
	\gamma^n_{mi}q^{ij}\gamma^m_{nj}+
	\gamma^m_{ij}q^{ij}\gamma^p_{pm}+
	\nonumber \\
	&&+\frac{M_p^2}{4}p_AK^{AB}p_B+\frac{1}{M_p^2}q^{ij}\partial_i
	\phi^A\partial_j\phi^B K_{AB} \ ,  
	\nonumber \\
	&&	\mC_i=-2\nabla_m (q^{mn}\mK_{in})+2\nabla_i(q^{mn}\mK_{mn}) \ . \  \nonumber \\
\end{eqnarray}
It is important to stress that  $q^{ij}$ is tensor density of order $2$ so that it is convenient
to introduce ordinary metric $h_{ij}$ that is related to $q_{ij}$ which is inverse to $q^{ij}$, by following prescription
\begin{equation}
	q_{ij}=M_p^2(\det q^{ij})^{-\frac{1}{2}}h_{ij} \ , 
\end{equation}
where the factor $M_p^2$ is inserted from dimensional reasons.
It can be shown 
that $h_{ij}$ 
 is metric compatible with connection $\gamma_{ij}^k$ \cite{Kluson:2025ajf,Kluson:2025uyl}
 \begin{equation}
	\partial_k h^{ij}+\gamma^i_{kr}h^{rj}+\gamma^j_{kr}h^{ri}=0 \ . 
\end{equation}
In other words $\gamma^i_{jk}$ are uniquely determined by $h_{ij}$ as 
\begin{equation}
	\gamma^i_{jk}=\frac{1}{2}h^{im}(\partial_j h_{mk}+\partial_k h_{mj}-\partial_m h_{jk}) \ .
\end{equation}
  Now in terms of canonical variables $h_{ij}$ and 
conjugate momenta $\pi^{ij}$ we obtain canonical form of the  action 
\cite{Kluson:2025ajf,Kluson:2025uyl}
\begin{eqnarray}\label{actfinal1}
	&&	S=\int d^4x \left(\partial_t h_{ij}\pi^{ij}+p_A\partial_t\phi^A+\Omega \tilde{\mC}+\Omega^i\tilde{\mC}_i+\right.\nonumber \\
	&&	+ M^2_{BI}  
	M_p^2\sqrt{h}\left(-\frac{N^2}{\Omega}+\frac{1}{\Omega}(\Omega^i+N^i)m_{ij}(\Omega^j+N^j)-
	\Omega h^{ij}m_{ij}\right)\left(1+\frac{1}{4M^2_{BI}M^2_p}V\right)-\nonumber \\
&&\left.-M^2_{BI}M^2_p \lambda N\sqrt{m}\right) \ , 
	\nonumber \\
\end{eqnarray}
where
\begin{eqnarray}
	&&	\tilde{\mC}=-4 M^2_{BI}M^2_p\sqrt{h}+\frac{1}{M_p^2\sqrt{h}}(\pi^{ij}h_{im}h_{jn}\pi^{mn}-\frac{1}{2}
	\pi^2)-M_p^2\sqrt{h}{}^{(3)}R +\nonumber \\
&&+\frac{1}{4 \sqrt{h}}p_AK^{AB}p_B+\sqrt{\det h}h^{ij}\partial_i\phi^A\partial_j\phi^B
K_{AB} \ , \nonumber \\
	&&	{}^{(3)}R= h^{ij}(-\partial_i\gamma^p_{pj}+
	\partial_m\gamma^m_{ij}-
	\gamma^n_{mi}\gamma^m_{nj}+
	\gamma^m_{ij}\gamma^p_{pm}) \ , \quad 
	\tilde{\mC}_i=-2\nabla_k (h_{ij}\pi^{jk})+p_A\partial_i\phi^A \ , \nonumber \\
\end{eqnarray}
and  where $\Omega$ and $\Omega^i$ are related to $\bPi^{00}$ and $\bPi^{0i}$ by following relations 
\begin{equation}
	\bPi^{00}=M_p^2\sqrt{h}\frac{1}{\Omega} \ , \quad  \bPi^{0i}=M_p^2\sqrt{h}\frac{\Omega^i}{ \Omega}\ .
\end{equation}
Finally we used 
 $3+1$ decomposition of metric $g_{\mu\nu}$
\cite{Dirac:1958sc,Arnowitt:1962hi} when we introduced   the lapse
function $N=1/\sqrt{-g^{00}}$ and the shift function
$N^i=-g^{0i}/g^{00}$. In terms of these variables we
write  components of the metric $g_{\mu\nu}$ as
\begin{eqnarray}
	g_{00}=-N^2+N_i m^{ij}N_j \ , \quad g_{0i}=N_i \ , \quad
	g_{ij}=m_{ij} \ ,
	\nonumber \\
	g^{00}=-\frac{1}{N^2} \ , \quad g^{0i}=\frac{N^i}{N^2} \
	, \quad g^{ij}=m^{ij}-\frac{N^i N^j}{N^2} \ ,
	\nonumber \\
\end{eqnarray}
where $m_{ij}$ is  three dimensional metric 
with inverse $m^{ij}$ and $m\equiv \det  m_{ij}$.

We observe that  the action (\ref{actfinal1}) still contains non-dynamical variables $N,N^i$ and $m^{ij}$.
In the spirit of Faddeev-Jackiw  approach \cite{Faddeev:1988qp}
 it is natural to eliminate them from the action 
by solving their equations of motion. Explicitly, from (\ref{actfinal1})
we obtain equation of motion for $N$ in the form 
\begin{eqnarray}
	-2\sqrt{h}\frac{N}{\Omega}\left(1+\frac{1}{4M^2_{BI}M^2_p}V\right)-\lambda M^2_{BI}M^2_p\sqrt{m}=0
	\nonumber \\
\end{eqnarray}
that can be solved for $N$ as 
\begin{equation}\label{Nsol}
N=-\frac{1}{2\sqrt{h}(1+\frac{1}{4M^2_{BI}M^2_p}V)}\Omega \lambda \sqrt{m} \ . 
\end{equation}
Further, equations of motion for $N^i$ implies
\begin{equation}
	\label{Nisol}
N^i=-\Omega^i \ .
\end{equation}
 Finally equations of motion for $m_{ij}$ that follow from 
(\ref{actfinal1}) have the form 
\begin{eqnarray}
	\sqrt{h}(\frac{1}{\Omega}(\Omega^i+N^i)(\Omega^j+N^j)-\Omega h^{ij})
	(1+\frac{1}{4M^2_{BI}M^2_p}V)-\frac{1}{2}\lambda N m^{ij}\sqrt{m}=0 \nonumber \\
\end{eqnarray}
that with the help of the  results (\ref{Nsol}) and (\ref{Nisol}) simplify considerably 
\begin{eqnarray}\label{meq}
	hh^{ij}(1+\frac{1}{4M^2_{BI}M^2_p}V)^2=\frac{1}{4}m^{ij}m \lambda^2 \ . 
\end{eqnarray}
This equation can be solved for  $m^{ij}$ as 
\begin{equation}
	  m^{ij}=\frac{\lambda}{2
	(1+\frac{1}{4M^2_{BI}M^2_p}V)}h^{ij} \ .
\end{equation}
Then inserting this result into (\ref{Nsol}) we finally get
\begin{equation}
	N^2=\frac{2}{\lambda}\Omega^2(1+\frac{1}{4M^2_{BI}M^2_p}V) \ . 
\end{equation}
Collecting all these results and inserting them into $\mC$ we obtain that scalar fields have following
contribution to $\tilde{\mC}$ 
\begin{eqnarray}
	\tilde{\mC}_{matt}=
\frac{1}{4\sqrt{h}}p_AK^{AB}p_B+\sqrt{h}h^{ij}
\partial_i\phi^A\partial_j\phi^BK_{AB}
-\frac{4}{\lambda}M^2_{BI}M^2_p\sqrt{h}(1+\frac{1}{4M^2_{BI}M^2_p}V)^2
\nonumber \\
\end{eqnarray}
that differs from the standard form of Hamiltonian constraint
by presence of term proportional to $V^2$. However it turns out
that it is possible to modify Born-Infeld inspired gravity action that reproduces standard form of Hamiltonian constraint for scalar field  \cite{Vollick:2005gc} as we explicitly demonstrate.
\subsection{Modified Scalar Field Contribution}
The modified scalar field contribution was proposed in 
\cite{Vollick:2005gc} and has the form 
\begin{eqnarray}\label{BIMact2}
	&&
	S_{BIM}=M_p^2M_{BI}^2\int d^4x [\sqrt{-\det \bA}-\lambda\sqrt{-\det(1+\frac{1}{4M^2_{BI}M^2_p}V)^{\frac{1}{2}}g_{\mu\nu}}] \ , \nonumber 
	\\
	&&
	\bA_{\mu\nu}=g_{\mu\nu}+\frac{1}{M^2_{BI}}(R_{(\mu\nu)}+\frac{1}{M^2_p}K_{AB}\partial_\mu\phi^A
	\partial_\nu\phi^B) \  .
	\nonumber \\
\end{eqnarray}
In this case the canonical action has the same form as in the  previous section with the exception that terms which depend on metric $g_{\mu\nu}$ give following contribution  
\begin{eqnarray}\label{auxterm}
&&	+M^2_{BI}  
	M_p^2\sqrt{h}\left(-\frac{N^2}{\Omega}+\frac{1}{\Omega}(\Omega^i+N^i)m_{ij}(\Omega^j+N^j)-
	\Omega h^{ij}m_{ij}\right)(1+\frac{1}{4M^2_{BI}M^2_p}V)-\nonumber \\
&&	-M^2_{BI}M^2_p \lambda N\sqrt{m}(1+\frac{1}{4M^2_{BI}M^2_p}V) \ . 
	\nonumber \\
\end{eqnarray}
Then it is easy to solve equations of motion that follow from (\ref{auxterm}) with the result 

\begin{equation}
	m^{ij}=\frac{\lambda^2}{2}h^{ij} \ , \quad N^2=\frac{2}{\lambda^2}\Omega^2 \ , \quad 
	N^i=-\Omega^i \ . \nonumber \\
\end{equation}
Then inserting this solution into (\ref{auxterm}) we obtain 
that  matter contribution to $\tilde{\mC}_{matt}$ has the form 
\begin{eqnarray}
	\tilde{\mC}_{matt}=
	\frac{1}{4\sqrt{h}}p_AK^{AB}p_B+\sqrt{h}h^{ij}
	\partial_i\phi^A\partial_j\phi^BK_{AB}
	-\frac{4}{\lambda^2}M^2_{BI}M^2_p\sqrt{h}-\frac{1}{\lambda^2}\sqrt{h}V 
	\nonumber \\
\end{eqnarray}
which corresponds to the standard matter contribution to the Hamiltonian constraint. In other words
Born-Infeld inspired gravity action (\ref{BIMact2})  is completely equivalent to the
General Relativity action coupled minimally to scalar fields. 

\section{Proposal of Eddington Gravity with Scalar Fields}\label{third}
Previous analysis suggests that in case of the matter content which is described by scalar
fields it is possible to have Born-Infeld inspired gravity whose canonical formulation exactly 
corresponds to General Relativity coupled minimally to scalar fields. It is remarkable that 
space-time metric appears as completely auxiliary field that is integrated out. Then we can 
ask the question whether it is possible to return to original Eddington's idea and formulate
action for gravity  and matter where dynamical degrees of freedom correspond to connection and 
scalar fields only. Motivated by previous analysis we propose such a form of action in the form 
\footnote{Similar action was previously found in \cite{Azri:2021dbr} however in different context.}
\begin{eqnarray}
	&&
	S_{EG\phi}=\int d^4x \frac{1}{V(\phi)}\sqrt{-\det \bA} \ , \nonumber 
	\\
	&&
	\bA_{\mu\nu}=R_{(\mu\nu)}+\partial_\mu\phi^A \partial_\nu \phi^B K_{AB}(\phi) \  , 
	\nonumber \\
\end{eqnarray}
where $V(\phi)$ and $K_{AB}(\phi)$ are functions of scalars fields where $V(\phi)=1+\tilde{V}(\phi), \tilde{V}(\phi=0)=0$. 

Let us find canonical form of this action. The procedure is almost the same as in previous section so that Hamiltonian density is equal to
\begin{eqnarray}
&&	\mH=-\bPi^{\mu\nu}
	(\partial_iG^i_{\mu\nu}+\frac{1}{3}G^\lambda_{\lambda \mu}G^\sigma_{\sigma\nu}-G^\lambda_{\sigma\mu}G^\sigma_{\lambda\nu})- 4V\sqrt{-\bPi}+
	\nonumber \\
&&	+\frac{1}{4\bPi^{00}}p_AK^{AB}p_B-\frac{\bPi^{i0}}{
		\bPi^{00}}p_A\partial_i\phi^A-
	\frac{1}{\bPi^{00}}(\bPi^{00}\bPi^{ij}-\bPi^{0i}\bPi^{0j})
	K_{AB}\partial_i\phi^A\partial_j\phi^B \ . 
	\nonumber \\
\end{eqnarray}
Then we can proceed exactly in the same way as in previous section with the final result
\begin{eqnarray}\label{actfinalEdd}
	&&	S=\int d^4x (\partial_t h_{ij}\pi^{ij}+p_A\partial_t\phi^A+\Omega \tilde{\mC}+\Omega^i\tilde{\mC}_i) \ , \nonumber \\
\end{eqnarray}
where
\begin{eqnarray}
	&&	\tilde{\mC}=\frac{1}{\sqrt{h}}(\pi^{ij}h_{im}h_{jn}\pi^{mn}-\frac{1}{2}
	\pi^2)-\sqrt{h}{}^{(3)}R +\nonumber \\
	&&+\frac{1}{4 \sqrt{h}}p_AK^{AB}p_B+\sqrt{\det h}h^{ij}\partial_i\phi^A\partial_j\phi^B
	K_{AB} -4V \sqrt{h}
		 \ , \nonumber \\
	&&
	\tilde{\mC}_i=-2\nabla_k (h_{ij}\pi^{jk})+p_A\partial_i\phi^A \ . \nonumber \\
\end{eqnarray}
This is final form of canonical action for Eddington gravity that couples to scalar field with potential $4V$. We see that it completely reproduces  General Relativity action coupled minimally to scalar field with potential $4V$. This remarkable achievement  explicitly demonstrates possibility  to formulate theory of gravity that couples to scalar fields where the dynamical degree of freedom corresponds to connection
while metric emerges as momentum conjugate to the specific combinations of connection.

{\bf Acknowledgment:}

This work  is supported by the grant “Dualitites and higher order derivatives” (GA23-06498S) from the Czech Science Foundation (GACR).


\begin{thebibliography}{20}




\bibitem{Dirac:1958sc}
P.~A.~M.~Dirac,
\emph{``The Theory of gravitation in Hamiltonian form,''}
Proc. Roy. Soc. Lond. A \textbf{246} (1958), 333-343
doi:10.1098/rspa.1958.0142

\bibitem{Arnowitt:1962hi}
R.~L.~Arnowitt, S.~Deser and C.~W.~Misner,
\emph{``The Dynamics of general relativity,''}
Gen. Rel. Grav. \textbf{40} (2008), 1997-2027
doi:10.1007/s10714-008-0661-1
[arXiv:gr-qc/0405109 [gr-qc]].




\bibitem{Gourgoulhon:2007ue}
E.~Gourgoulhon,
\emph{``3+1 formalism and bases of numerical relativity,''}
[arXiv:gr-qc/0703035 [gr-qc]].





\bibitem{Horava:1990ba}
P.~Horava,
\emph{``On a covariant Hamilton-Jacobi framework for the Einstein-Maxwell theory,''}
Class. Quant. Grav. \textbf{8} (1991), 2069-2084
doi:10.1088/0264-9381/8/11/016



\bibitem{Dirac:1958sc}
P.~A.~M.~Dirac,
\emph{``The Theory of gravitation in Hamiltonian form,''}
Proc. Roy. Soc. Lond. A \textbf{246} (1958), 333-343
doi:10.1098/rspa.1958.0142
%

\bibitem{Faddeev:1982id}
L.~D.~Faddeev,
\emph{``The energy problem in Einstein's theory of gravitation,''}
Sov. Phys. Usp. \textbf{25} (1982), 130-142
doi:10.1070/PU1982v025n03ABEH004517

\bibitem{McKeon:2010nf}
D.~G.~C.~McKeon,
\emph{``The Canonical Structure of the First Order Einstein-Hilbert Action,''}
Int. J. Mod. Phys. A \textbf{25} (2010), 3453-3480
doi:10.1142/S0217751X10050093
[arXiv:1005.3001 [gr-qc]].


\bibitem{Chishtie:2013fna}
F.~Chishtie and D.~G.~C.~McKeon,
\emph{``The Canonical Structure of the First Order Einstein-Hilbert Action with a Flat Background,''}
Class. Quant. Grav. \textbf{30} (2013), 155002
doi:10.1088/0264-9381/30/15/155002
[arXiv:1304.5613 [gr-qc]].

\bibitem{Kiriushcheva:2011aa}
N.~Kiriushcheva, S.~V.~Kuzmin and D.~G.~C.~McKeon,
\emph{``An Analysis of the First Order Form of Gauge Theories,''}
Can. J. Phys. \textbf{90} (2012), 165-174
doi:10.1139/p11-154
[arXiv:1112.2003 [hep-th]].

\bibitem{Kiriushcheva:2010ycc}
N.~Kiriushcheva and S.~V.~Kuzmin,
\emph{``The Hamiltonian formulation of General Relativity: Myths and reality,''}
Central Eur. J. Phys. \textbf{9} (2011), 576-615
doi:10.2478/s11534-010-0072-2
[arXiv:0809.0097 [gr-qc]].

\bibitem{Kiriushcheva:2010pia}
N.~Kiriushcheva and S.~V.~Kuzmin,
\emph{``The Hamiltonian of Einstein affine-metric formulation of General Relativity,''}
Eur. Phys. J. C \textbf{70} (2010), 389-422
doi:10.1140/epjc/s10052-010-1458-4
[arXiv:0912.3396 [gr-qc]].

\bibitem{Kiriushcheva:2006gp}
N.~Kiriushcheva, S.~V.~Kuzmin and D.~G.~C.~McKeon,
\emph{``A Canonical Analysis of the Einstein-Hilbert Action in First Order Form,''}
Int. J. Mod. Phys. A \textbf{21} (2006), 3401-3420
doi:10.1142/S0217751X06029545
[arXiv:hep-th/0609219 [hep-th]].





\bibitem{Faddeev:1988qp}
L.~D.~Faddeev and R.~Jackiw,
\emph{``Hamiltonian Reduction of Unconstrained and Constrained Systems,''}
Phys. Rev. Lett. \textbf{60} (1988), 1692-1694
doi:10.1103/PhysRevLett.60.1692




\bibitem{Jackiw:1993in}
R.~Jackiw,
\emph{``(Constrained) quantization without tears,''}
[arXiv:hep-th/9306075 [hep-th]].





\bibitem{Garcia:1996ac}
J.~A.~Garcia and J.~M.~Pons,
\emph{``Equivalence of Faddeev-Jackiw and Dirac approaches for gauge theories,''}
Int. J. Mod. Phys. A \textbf{12} (1997), 451-464
doi:10.1142/S0217751X97000505
[arXiv:hep-th/9610067 [hep-th]].




\bibitem{Kluson:2025ajf}
J.~Kluson,
\emph{``Canonical Analysis of Eddington Gravity,''}
[arXiv:2506.16279 [gr-qc]].

\bibitem{Kluson:2025uyl}
J.~Kluson,
\emph{``Canonical Form of Born-Infeld Inspired Gravity Coupled to Scalar Fields,''}
[arXiv:2507.23424 [gr-qc]].

\bibitem{Vollick:2005gc}
D.~N.~Vollick,
\emph{``Born-Infeld-Einstein theory with matter,''}
Phys. Rev. D \textbf{72} (2005), 084026
doi:10.1103/PhysRevD.72.084026
[arXiv:gr-qc/0506091 [gr-qc]].



\bibitem{Kibaroglu:2024ico}
S.~Kibaro{\u{g}}lu, S.~D.~Odintsov and T.~Paul,
\emph{``Cosmology of unimodular Born{\textendash}Infeld-fR gravity,''}
Phys. Dark Univ. \textbf{44} (2024), 101445
doi:10.1016/j.dark.2024.101445
[arXiv:2402.08951 [gr-qc]].

\bibitem{Makarenko:2014lxa}
A.~N.~Makarenko, S.~Odintsov and G.~J.~Olmo,
\emph{``Born-Infeld-$f(R)$ gravity,''}
Phys. Rev. D \textbf{90} (2014), 024066
doi:10.1103/PhysRevD.90.024066
[arXiv:1403.7409 [hep-th]].

\bibitem{Azri:2021dbr}
H.~Azri and S.~Nasri,
\emph{``Dynamical aspects of asymmetric Eddington gravity with scalar fields,''}
Phys. Rev. D \textbf{104} (2021) no.6, 064028
doi:10.1103/PhysRevD.104.064028
[arXiv:2107.12953 [gr-qc]].


\bibitem{BeltranJimenez:2017doy}
J.~Beltran Jimenez, L.~Heisenberg, G.~J.~Olmo and D.~Rubiera-Garcia,
\emph{``Born{\textendash}Infeld inspired modifications of gravity,''}
Phys. Rept. \textbf{727} (2018), 1-129
doi:10.1016/j.physrep.2017.11.001
[arXiv:1704.03351 [gr-qc]].



\bibitem{Padmanabhan:2013nxa}
T.~Padmanabhan,
\emph{``General Relativity from a Thermodynamic Perspective,''}
Gen. Rel. Grav. \textbf{46} (2014), 1673
doi:10.1007/s10714-014-1673-7
[arXiv:1312.3253 [gr-qc]].


\bibitem{Eddington}
A. Eddington, \emph{"The Mathematical Theory of Relativity,"} Cambridge University Press, Cambridge,
UK, 2 ed., 1924.






\end{thebibliography}
\end{document}